# ИСПОЛЬЗОВАНИЕ МЕТОДОВ ИСКУССТВЕННОГО ИНТЕЛЛЕКТА ДЛЯ ИССЛЕДОВАНИЙ ЗРИТЕЛЬНОГО АНАЛИЗАТОРА[1]

А. И. Медведева (*Medvedeva.AI@rea.ru*)
М.В. Холод (*Kholod.MV@rea.ru*)
Российский экономический университет имени Г.В. Плеханова, Москва

В работе описываются способы использования различных методик применения искусственного интеллекта для исследования глаз человека. Первый дата сет был собран при помощи компьютерной периметрии для исследования визуализации поля зрения человека и постановки диагноза глаукома. Предлагается способ анализа изображения при помощи программных средств. Второй датасет был получен в рамках реализации российско-швейцарского эксперимента по сбору и анализу данных о движении глаз при помощи прибора «Tobii Pro Glasses 3» на VR-видео. Были исследованы движения и фокусировка глаз на записанном маршруте виртуального путешествия по кантону Во. Разрабатываются методики исследований зависимостей движений зрачка глаза при помощи математического моделирования. Данные исследования могут быть использованы в медицине для оценки течения и ухудшения глаукомы пациентов, а также для изучения механизмов внимания к туристическим достопримечательностям пользователей VR-видео.

**Ключевые слова**: глаукома, периметрия, слепое пятно, айтрекинг «Tobii Pro Glasses 3», VR-video, искусственный интеллект.

## Введение

Зрительный анализатор является одним из важнейших органов чувств человека, обеспечивающим восприятие и обработку визуальной информации. Исследование зрительного анализатора имеет большое значение для понимания механизмов восприятия, диагностики и

---

лечения заболеваний глаз, а также разработки новых технологий в области офтальмологии.

В последние годы методы искусственного интеллекта (ИИ) получили широкое применение в различных областях медицины, включая офтальмологию. ИИ-алгоритмы способны анализировать большие объемы данных, выявлять закономерности и принимать решения, что открывает новые возможности для исследований зрительного анализатора.

В данной статье рассмотрены различные методы ИИ, используемые для исследования зрительного анализатора, их преимущества и ограничения. Были изучены перспективы применения ИИ в офтальмологии и его потенциальное влияние на диагностику и лечение заболеваний глаз.

Глаукома – это заболевание зрительного нерва, которое может привести к оптической невропатии вплоть до полной потери зрения. Слепое пятно – это участок глаза, в котором отсутствуют светочувствительные клетки (палочки и колбочки). Одним из способов диагностики заболеваний глаукомы на основе изучения полей зрения является периметрия.

Существует два вида периметрии: статическая и кинетическая. Статическая периметрия основана на использовании неподвижного объекта при исследовании поля зрения пациента. При этом применяется различная освещённость объекта, и информация передается на ПК. Этот тип анализа поля зрения пациента в конечном счёте приводит к получению результата в виде картограмм, которые подвергаются обработке. При кинетической периметрии используют объект разных цветов, который находится в движении [Сомов, 2016].

«Tobii Pro Glasses 3» это портативный айтрекер третьего поколения, разработанный для исследований в реальном времени и в самых разных условиях. Увеличенное поле зрения камеры сцены в очках большое значение при проведении исследований на открытом воздухе, где из-за более узкого поля зрения теряются данные о взгляде.

Окуломоторная активность является необходимым компонентом психических процессов, связанных с получением, преобразованием и использованием зрительной информации, а также состояний, деятельности и общения человека. Поэтому, регистрируя и анализируя движения глаз, исследователь получает доступ к скрытым (внутренним) формам активности, которые обычно протекают в свернутой форме, исключительно быстро и неосознанно. Как показывают исследования, по характеру движений глаз можно определить:

направленность взора, стратегии прослеживания движущихся объектов, маршруты сканирования воспринимаемых сцен, информационную сложность объекта и т.д. [Ананьева и др., 2015].

Целью работы является компьютерный анализ снимков-картограмм при помощи программных средств после прохождения периметрии для исследования развития диагноза глаукома. А также использование методов кластеризации и тепловых карт при обработке информации, полученной путём айтрекинга и влияние взгляда на восприятие изображений и их оценку в головном мозге.

# 1. Исследования поля зрения пациентов при помощи компьютерного анализа изображения

## 1.1. Исследование анатомии глаза

При изучении поля зрения пациентов в начале следует рассмотреть глазное яблоко, в котором находятся рецепторы зрительного анализатора. Данный орган расположен приблизительно на две трети в полости глазницы, заполненной в заднем отделе жировым телом. Через ее пространство проходят также соединительная оболочка, образующая глазодвигательные мышцы, кровеносные сосуды, ветви ряда двигательных и чувствительных нервов [Сомов, 2016].

Способность глаза реагировать на возможно малый поток излучения называется световой чувствительностью. Реакция глаза зависит от потока излучения, упавшего на сетчатку и от той доли спектральной мощности, которая попадает на рецепторы.

Природа заложила в каждом глазу (за исключением патологий) критерий равный «1», который имеет название «относительной спектральной световой эффективностью». От данного критерия отсчитывают весь уровень световых потоков – красного, синего, жёлтого и др. [ Сухинин и др., 2012].

## 1.2. Разработка алгоритма по анализу поля зрения пациентов

Предполагается, что зона слепого пятна находится в височной стороне примерно в 12°–15° от центра, чуть ниже горизонтальной оси. Вначале задаются минимальные границы, где слепое пятно должно находиться анатомически, это примерно 10° – 18° по полярному радиусу, а для правого глаза (-20° – 10°). Также возможно определить зоны, которые прилегают к слепому пятну, в них присутствует абсолютная скотома. Наличие абсолютной скотомы в этих зонах говорит о значительном повреждении сетчатки и невозможности провести кинетическую периметрию слепого пятна в автоматическом режиме. Алгоритм анализа снимков основан на компьютерной обработке зарегистрированных положений стимулов, пола, возраста и диагноза пациента. В результате чего получается компьютерное изображение с положением точек, на основе которого в дальнейшем возможно создание классификатора диагноза испытуемого при большом количестве датасета, а также создания отечественного ПО для Российского периметра.

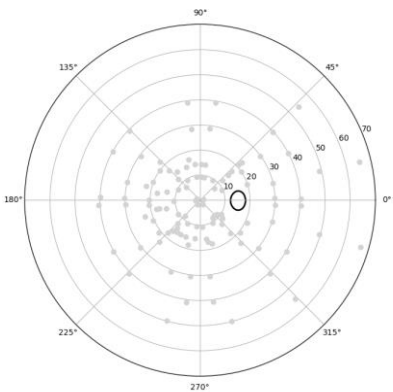

Рис.1. Результат обработки картограммы правого глаза пациента, больного глаукомой, смоделированный в Python на основании снимка его периметрии

## 2. Исследование движения глаз зрителей VR-видео при помощи технологии айтрекинга

### 2.1. Исследование и сбор данных движения глаз зрителей VR-видео

Один из методов, который привлекает значительное внимание в современных исследователей виртуальной реальности, предполагает использование данных, собранных с сенсорного оборудования, которое носит пользователь, в качестве способа анализа различных аспектов его состояния или поведения во время восприятия виртуальной реальности. Например, изучение изменений движения глаз пользователя (точка фокусировки, движение взгляда, диаметр зрачка) на основе таких данных, собранных с устройства слежения за глазами возможно увидеть взаимосвязь между такими факторами, как стресс, вовлеченность, удовлетворенность и качество разговора [Jeong, 2022].

В январе 2023 года был проведен эксперимент по сбору данных о движениях глаз с участием 16 испытуемых. Целью эксперимента было предложить участникам просмотреть видео в формате 360°, отслеживая и записывая движения глаз и положения зрачков с помощью очков Tobii Pro Glasses 3. Эти очки могут записывать со скоростью 50 или 100 кадров в секунду, а технология автоматически подстраивается под саккады для обеспечения точности измерения данных. Видео 360° было создано в Бюро туризма Кантона Во в Швейцарии.

Значительное преимущество этой технологии заключается в устранении предвзятости или отношения участников к объекту исследования в исследованиях с отслеживанием взгляда [Duchowski, A. T., 2000]. Следовательно, точность и достоверность полученных данных остаются неизменными, не подвергаясь влиянию мотивации участников во время исследования.

## 2.2. Описание данных движения глаз зрителей VR-видео по сценам

В ходе эксперимента собиралась следующая информация: номер события (одинаково для 1 человека); абсолютное время начала события (микросекунды), абсолютное время окончания события (микросекунды); точка взгляда X как координата X (пиксели экрана), точка взгляда Y как координата Y (пиксели экрана); отклонение по X (пикселям экрана), отклонение по Y (пикселям экрана); точка взгляда 3D X, точка взгляда 3D Y, точка взгляда 3D Z; направление взгляда влево/вправо X, направление взгляда влево/вправо Y, направление взгляда влево/вправо Z; положение зрачка слева/справа X, положение зрачка слева/справа Y, положение зрачка слева/справа Z; диаметр зрачка слева (справа), диаметр отфильтрованного зрачка; тип движения глаз; продолжительность события взгляда (микросекунды); индекс типа движения глаз; точка фиксации X, точка фиксации Y; гироскоп X, гироскоп Y, гироскоп Z; акселерометр X, акселерометр Y, акселерометр Z.

Для целей исследования были использованы данные о количестве событий (одинаково для 1 человека); абсолютное время начала события (микросекунды), абсолютное время окончания события (микросекунды); точка взгляда X как координата X (пиксели экрана), точка взгляда Y как координата Y (пиксели экрана).

Во-первых, для того, чтобы разбить видео на сцены, используется абсолютное время начала события, поэтому было разделено видео 360° на 15 сцен – Валле-де-Жу, Монтрё-Рош-де-Нэ, Лозанна (Собор), Лозанна (Уши), Морж, Ньон, Монтре (Шильонский замок), Лаво-ле-Бен, Жюра-Водуа, Аванш, Эгль, Шато-д'О, Лейзен (Куклос), Ле-Дьяблере (Ледник 3000).

Абсолютная точка взгляда X как координата X (пиксели экрана) и точка взгляда Y как координата Y (пиксели экрана) используются для визуализации движений глаз зрителя (рис.2 и рис.3).

Проблема такой визуализации что, если сцены изменят порядок, измерения точек взгляда Xs как координат X (пиксели экрана) и точек взгляда Y как координаты Y (пиксели экрана) также изменятся [Wu, X., 2022].

Во-вторых, было рассчитано описательную статистику траекторного пути. Из таблицы 1 видно, что наиболее привлекательная для зрителя сцена – Ле Дьяблере (Ледник 3000) с числом точек фиксации 2105 и количеством точек саккад 154, вторая привлекательная для зрителя сцена – Монтрё (Рош-де-Нэ) с количество точек фиксации 1776 и количество точек саккад 106, следующие – Лозанна, Лейзен (Куклос), Лозанна (Собор) и Монтре (ривьера) привлекли наибольшее внимание зрителя, а наиболее активные саккады были на сценах ледника и Монтреа (ривьера). Частоты дают определенные характеристики туристических предпочтений зрителя.

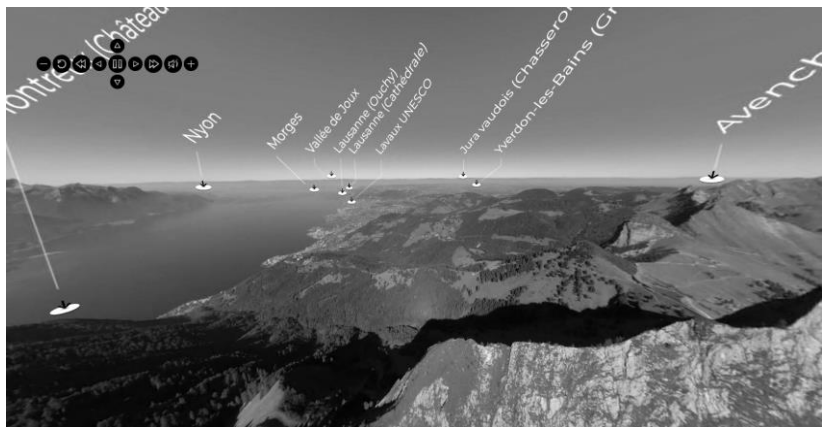

Рис.2. Структура VR видео 360° кантона Во (Валле-де-Жу, Монтрё (Рош-де-Нэ), Лозанна (Собор), Лозанна (Уши), Морж, Ньон, Монтрё (Шильонский замок), Лаво-ле- Бен, Жюра Водуа, Аванш, Эгль, Шато-д'О, Лейзен (Куклос), Ле-Дьяблере (Ледник 3000)

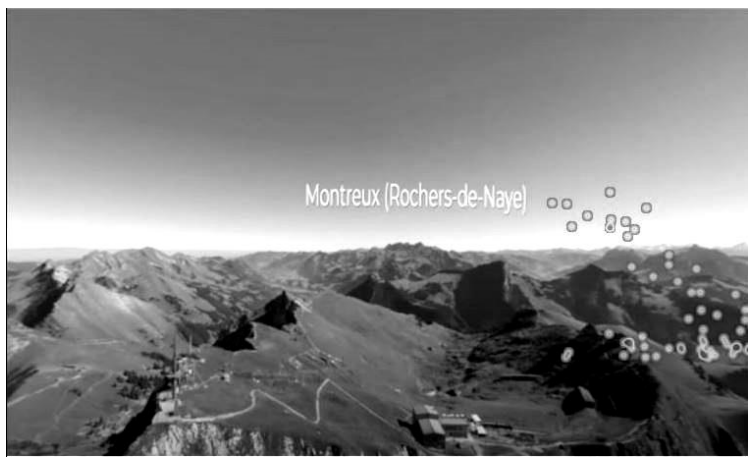

Рис.3. Пример тепловой карты, построенной на примере сцены Монтрё

(Рош-де-Нэ)

### 2.3. Описание траекторий движения глаз по сценам

Описание разброса координат приведено в таблице 1, похожая аналитика приведена в статье [Almquist, M., 2017]. Видно, что самый широкий разброс в движении глаз характерен для широких природных сцен, таких как Валле-де-Жу, Монтрё (Рош-де-Нэ), Ньон, Монтрё (Шильонский замок), Лаво-ЮНЕСКО, Аванш, Лейзен, Ле Дьябле.

Табл.1

| Сцена | Фиксация | Саккада | Неклассифицированный |
|---|---|---|---|
| Эгль | 856 | 42 | |
| Аванш | 862 | 50 | 1 |
| Шато-д'О | 785 | 22 | |
| Жура Водуа (Шассерон) | 560 | 36 | |
| Лозанна (Собор) | 1435 | 70 | 1 |
| Лозанна (Уши) | 1462 | 56 | |
| Лаво-ЮНЕСКО | 1187 | 50 | |
| Ле Дьяблере (Ледник 3000) | 2105 | 154 | 3 |
| Лейзин (Куклос) | 1125 | 52 | 5 |
| Монтре (Шильонский замок) | 508 | 28 | |
| Монтрё (Рош-де-Нэ) | 1776 | 106 | 1 |
| Морж | 760 | 48 | |
| Ньон | 609 | 57 | |
| Вали де Жу | 689 | 46 | |
| Ивердон-ле-Бен | 496 | 41 | |

Результаты, основанные на приведенном выше анализе, заключаются в том, что (1) движения глаз зрителей более широки в естественных и широких сценах, (2) движения глаз зрителя более интенсивны в сценах с некоторыми объектами, например, дом в горах, корабль на озере, собор в городе.

### Заключение

Искусственный интеллект (ИИ) произвел революцию в области исследований зрительного анализатора, открыв новые возможности для понимания механизмов восприятия, диагностики и лечения заболеваний глаз. Методы ИИ, такие как машинное обучение, глубокое обучение и компьютерная обработка изображений, позволили анализировать большие объемы данных, выявлять сложные закономерности и принимать обоснованные решения.

В данной статье рассмотрены различные возможности программных средств, используемые для исследования зрительного анализатора, их преимущества и ограничения. Проведено обсуждение, как ИИ может улучшить диагностику заболеваний, таких как глаукома, путем анализа полей зрения и данных о движениях глаз. Был изучен потенциал ИИ в разработке новых методов лечения, персонализированных для индивидуальных пациентов.

Будущие исследования должны сосредоточиться на разработке более точных и надежных методов ИИ, а также на интеграции ИИ в клиническую практику, например, создании ПО для автоматического анализа снимков или различных IT- решений в области использования айтрекинга в экономике и туризме. В части исследования глаукомы необходимо рассмотреть возможность обучения нейронных сетей на более крупном датасете.

Сотрудничество между исследователями в области ИИ, офтальмологами и другими специалистами в области здравоохранения имеет решающее значение для реализации полного потенциала ИИ в области исследований зрительного анализатора.

Вклад авторов: проведение эксперимента в России и Швейцарии по айтрекингу, разработка дизайна этого исследования, помимо данных айтрекинга собраны анкетные данные. Основной массив данных находится в процессе обработки. С помощью метода тепловых карт проведена частичная визуализация данных. Сбор датасета компьютерной периметрии, отбор снимков для компьютерного анализа, создание базы данных обработанных снимков, написание норм и критериев создания и передачи снимков для врачей. Возможно дальнейшее создание классификатора, определяющего норму/не норму. Для этого необходимо дальнейшее увеличение дата сета.

## Список литературы

# USING ARTIFICIAL INTELLIGENCE METHODS FOR THE STUDIED VISUAL ANALYZER

A. I. Medvedeva (Medvedeva.AI@rea.ru)
M.V. Kholod (Kholod.MV@rea.ru)
G.V. Plekhanov Russian University of Economics, Moscow

The paper describes how various techniques for applying artificial intelligence to the study of human eyes are utilized. The first dataset was collected using computerized perimetry to investigate the visualization of the human visual field and the diagnosis of glaucoma. A method to analyze the image using software tools is proposed. The second dataset was obtained, as part of the implementation of a Russian-Swiss experiment to collect and analyze eye movement data using the Tobii Pro Glasses 3 device on VR video. Eye movements and focus on the recorded route of a virtual journey through the canton of Vaud were investigated. Methods are being developed to investigate the dependencies of eye pupil movements using mathematical modelling. VR-video users can use these studies in medicine to assess the course and deterioration of glaucoma patients and to study the mechanisms of attention to tourist attractions.

**Keywords**: glaucoma, perimetry, blind spot, "Tobii Pro Glasses 3", VR-video, artificial intelligence.